# What X-ray absorption spectroscopy can tell us about the active state of earth-abundant electrocatalysts for the oxygen evolution reaction

Marcel Risch,*[a] Dulce M. Morales,[a] Javier Villalobos[a] and Denis Antipin[a]

*Nachwuchsgruppe Gestaltung des Sauerstoffentwicklungsmechanismus, Helmholtz-Zentrum Berlin für Materialien und Energie GmbH, Hahn-Meitner-Platz 1, Berlin 14109, Germany.*
*\* marcel.risch@helmholtz-berlin.de.*

Chemical energy storage is an attractive solution to secure a sustainable energy supply. It requires an electrocatalyst to be implemented efficiently. In order to rationally improve the electrocatalyst materials and thereby the reaction efficiency, one must reveal the nature of the electrocatalyst under reaction conditions, i.e., its active state. For a better understanding of earth-abundant metal oxides as electrocatalysts for the oxygen evolution reaction (OER), the combination of electrochemical (EC) methods and X-ray absorption spectroscopy (XAS) has been very insightful and still holds untapped potential. Herein, we concisely introduce the basics of EC and XAS and provide the necessary framework to discuss changes that electrocatalytic materials undergo, presenting manganese oxides as examples. Such changes may occur during preparation and storage, during immersion in an electrolyte, as well as during application of potentials without or with catalytic reactions. We conclude with a concise summary of how EC and XAS are currently combined to elucidate the active state as well as an outlook on future opportunities to understand the mechanisms of electrocatalysis using combined operando EC-XAS experiments.

## Introduction

A sustainable energy supply is of global interest and one of the grand societal challenges of the century.[1,2] The UN defined it as one of its 17 goals for sustainable development, namely, to "ensure access to affordable, reliable, sustainable and modern energy for all".[3] One crucial target is to increase the use of renewable energy, which necessitates storage to buffer the intermittence of renewable sources such as wind and sun. Chemical energy storage is an attractive solution. It requires an electrocatalyst that provides reaction intermediates to lower the energy barriers, thereby increasing the storage efficiency. In order to rationally improve electrocatalyst materials and thus the reaction efficiency, the nature of the electrocatalyst under reaction conditions must be revealed.[4,5] Unfortunately, many important details of the active states of electrocatalysts under reaction conditions are still unknown.[6,7]

For a better understanding of electrocatalysts, the combination of electrochemical (EC) methods and X-ray absorption spectroscopy (XAS) has been very insightful. The key advantage of XAS over spectroscopic and diffraction methods is its element specifity, which allows correlating EC processes to oxidation state changes of specific elements and which further allows elucidating the changes in the coordination environment of these elements. For the latter, no long-range order is required. Thus combined EC and XAS experiments hold much untapped potential to understand EC processes. Several aspects of combined EC and XAS experiments topic have been reviewed in the last decade approached from the perspective of the synchrotron science community.[8–11] Here, we will complement the prior work by approaching the topic and its synergetic aspects from the perspective of an electrochemist.

What is measured in electrochemical experiments? In these experiments, electric current and voltages (defined as the difference between electrode potentials) are measured, where the former is the flow of electrons and the latter quantifies the capacity of the electrochemical system to do work. Common pitfalls in the physical interpretation of typical electrode potentials used in electrochemistry were recently discussed by Boettcher et al.[12] The application of an electrode potential may lead to various effects on a electrocatalyst material. For instance, it could induce a flow of electrons with only little change to it, e.g., metallic conduction via an electron gas. It may induce a chemical change in the material, such as a change in oxidation state of the metals or, in some cases, of the ligand oxygen. The latter changes may lead to complete transformation of the electrocatalyst, often as corrosion. The potential-induced changes may take place on the surface (defined here as the liquid-solid interface, including inner pores), near the surface or in the bulk.

What is measured by XAS? In these experiments, the X-ray absorption near edge structure (XANES) correlates with the number of electrons in the outermost shell and the extended X-ray absorption fine structure (EXAFS) relates to the local coordination environment of the absorbing element. A higher energy of the X-ray photon results in a larger penetration depth in the material and a larger escape depth being relevant for proxies of the X-ray absorption measured, e.g., by the X-ray fluorescence. For the materials and absorption edges covered herein, these depths (calculation in Supporting Information) are of the order from 0.3 μm (soft XAS, typically <1 keV ) to 60 μm (tender/hard XAS, typically >5 keV) for simple Mn oxides. Since nanosized materials are attractive electrocatalysts due to their high surface to bulk ratio, the XAS discussed herein probes, with some exceptions discussed later, the entire electrocatalyst material.

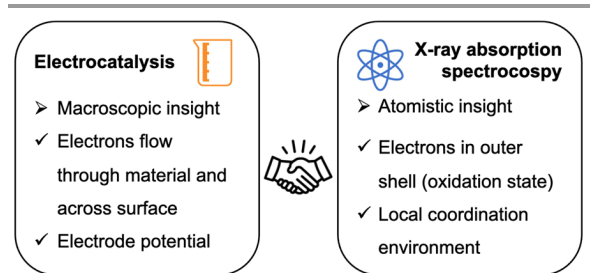

**Figure 1.** Key insight from X-ray absorption spectroscopy and electrochemistry.

Figure 1 compares the insights that can be gained from combined EC and XAS experiments, which offer several synergies. Electrochemistry provides a macroscopic insight into the behavior of the electrochemical cell and its components, where the working electrode is host to the catalyst under investigation and thus the most interesting. This is complemented by an atomistic insight into the electrocatalyst by XAS. The most valuable synergy is given by the ability to follow electrons through a combination of EC and XAS, where EC measures the flow rate of electrons, i.e., a current, while XAS measures the oxidation state, that is, the number of electrons in the outer shells, which determines the reactivity of the elements in the electrocatalyst. This is particularly insightful when coupled with additional electrochemical or non-electrochemical detection of the product of interest so that it becomes possible to correlate the current due to product generation with an oxidation state change, thus removing ambiguities about the origin of the electrons used for the formation of the desired product.

In this Feature Article, we concisely introduce the basics of EC and XAS, providing the necessary framework to discuss changes of the electrocatalyst material. As a way of example, we focus on manganese oxides as electrocatalysts for the oxygen evolution reaction (OER), which is the anodic reaction in water electrolyzers and currently considered the efficiency-limiting process in these devices.[4] We emphasize the importance of considering that changes may occur not only during a catalytic reaction, but also during preparation, storage, immersion in an electrolyte, or when applying potentials even if those do not lead to catalytic reactions. We conclude with a concise summary of how EC and XAS are currently combined to elucidate the active state of an electrocatalyst as well as an outlook on future opportunities for an in-depth understanding of the mechanisms of electrocatalysis.

## Brief fundamentals of electrochemistry

Electrochemistry relates chemical reactions to moving charges in response to gradients in the electric or chemical potential.[13] Any discontinuity between two materials may produce opposing charges at the interface; the (electrical) double layer, where charge transfer reactions take place. Within the scope of this Feature Article, we will only discuss the double layer in the liquid near a solid-liquid interface. When the potential difference is larger than the free energy of a given reaction, charge transfer becomes possible and ions, either in the liquid near the electrode surface or in the solid, may change their oxidation state by a redox reaction. In order for charge transfer to occur, the electrons need to be given sufficient energy to overcome kinetic and thermodynamic barriers between different states of the electrocatalysts, e.g., the barrier for Mn oxidation. A simple model for the relation between the diffnce in electrode potential and reaction rate (or current density) of a simple redox reaction is given by the Butler-Volmer equation.[14] Yet, catalytic reactions typically require multiple charge transfers of different types, e.g., the OER needs the transfer of four electrons and protons (acid) or hydroxide ions (base). The half reactions of the OER are:

Acid     $2H_2O \rightarrow O_2 + 4e^- + 4H^+$     (Eq.1)
Base     $4OH^- \rightarrow O_2 + 4e^- + 2H_2O.$     (Eq. 2)

Several proposals exist for possible intermediates.[15] Often the assumption is made that the reaction rate of one of the intermediates is significantly slower than those of the other steps, which defines the rate-limiting step and reduces the complex multi-charge transfer reaction to a simpler reaction with a single charge transfer.[16] The state prior to the slow step is also the most likely to be resolved by XAS (and other complementary methods).

Cyclic voltammetry (CV) is a convenient electrochemical method to probe and possibly distinguish processes with charge transfer (Faradaic processes) and those without charge transfer (non-Faradaic processes).[17] In this method, the electrode potential is swept back and forth within a potential window and the current is measured. By convention, sweeping towards more positive potentials means sweeping in the anodic direction and sweeping toward more negative potentials corresponds to the cathodic direction. Thereby, the method combines a thermodynamic property, namely potential, with kinetics measured as an electric current. Note that for the study of half reactions such as the OER, one often uses a third electrode to sense the potential near the electrode of interest (working electrode) and lets the current flow between the latter and an auxiliary electrode (also called counter electrode).The advantage of this three-electrode setup is that the measured current can be attributed fully to the reaction at the working electrode.

Figure 2 shows a typical CV experiment on an electrodeposited manganese oxide related to birnessite.[18] In this example, the potential region highlighted in blue in Figure 2a shows the fingerprint of a (quasi) reversible redox reaction on an oxide in a liquid, in this case, the redox transition from $Mn^{2+}$ to $Mn^{3+}$ in the anodic direction, and the corresponding reduction of $Mn^{3+}$ to $Mn^{2+}$ in the cathodic direction (Figure 2b, Table 1). Possible structural changes are schematically shown for the first coordination shell, where a Jahn-Teller distortion could occur due to oxidation of $Mn^{2+}$ ($d^5$ high spin) to localized $Mn^{3+}$ ($d^4$ high spin), thus leading to a change in apical bond lengths. Jahn-Teller distortions have been well studied, e.g., for $LiMn_2O_4$,[19] where it is well observed at temperatures below room temperature but not above.[19–22]

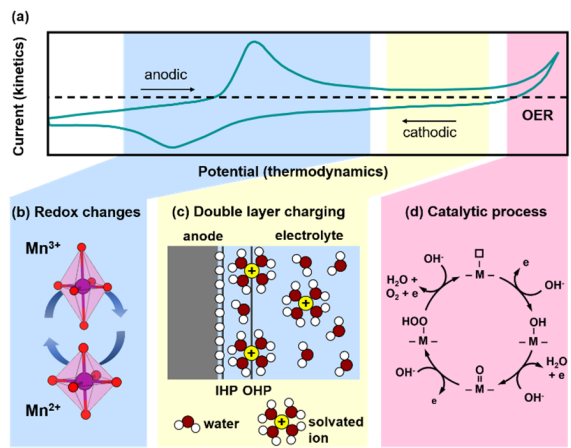

**Figure 2.** Anatomy of a cyclic voltammetry trace and associated physical processes. (a) cyclic voltammogram of electrodeposited MnOx and assignment of the electrochemical processes of (b) transition metal redox changes that may change the coordination environment, (c) double layer charging, and (d) electrocatalysis, here OER. Panel a was modified from ref.[18] with permission from the ACS. Panel c was modified from ref.[14] with permission from Wiley.and Sons. Panel d was reproduced from ref.[23] under a CC BY NC ND 4.0 License.

The potential region indicated with a yellow background in Figure 2a shows a capacitive behaviour, i.e., an increase in potential results in an unchanging current due to constant charge accumulation in the double layer that can be modeled as a plate capacitor (Figure 2c). Therefore, the charge increases with area and the data obtained in this region can be used to determine the area of the material in contact with the electrolyte so that the electrochemical surface area (ECSA) can be estimated (Table 1) when no Faradaic processes occur.[24] It is a common method but we[25] and others[24,26–29] have pointed out common pitfalls in its determination by voltammetric methods.

In the high potential region indicated in pink background in Figure 2a, there is an exponential increase in current in the anodic direction which, however, does not display a corresponding cathodic peak. This is a typical fingerprint of an irreversible reaction, here the OER, under the applied conditions. Several processes may contribute to the measured currents in addition to catalysis, namely the aforementioned Faradaic and non-Faradaic processes. After collection of the catalytic currents, either or both suitably normalized currents (i) and their corresponding electrode potentials (E) are used for determination of the electrocatalyst's activity and stability for benchmarking or for determiination of mechanistic parameters such as the Tafel slope, reaction order and Nernst slope (Table 1).[28,30–35] Recommendations and protocols for electrocatalyst benchmarking, including for example suitable correction and/or appropriate conditioning procedures, have been published by several groups;[28,30,31,35–37] a harmonization of the benchmarking endeavours is desirable but has not been agreed upon yet.

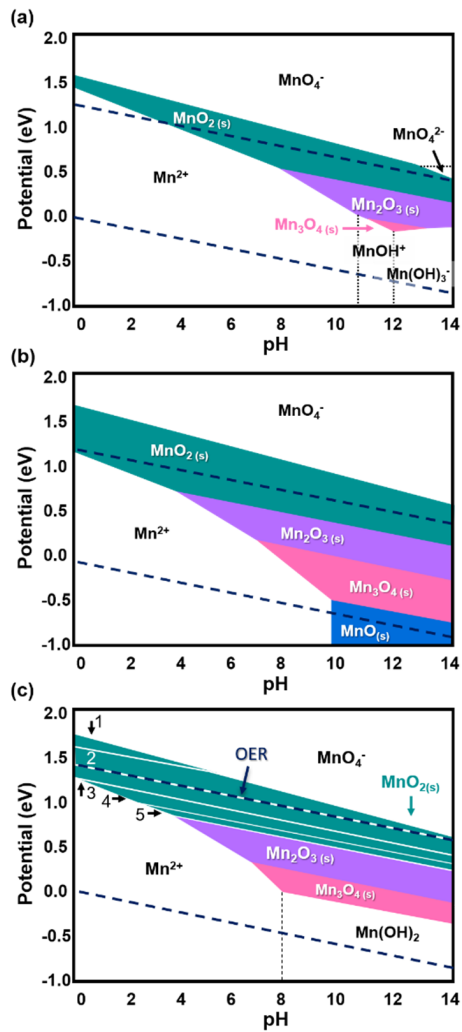

**Figure 3.** E-pH (Pourbaix) diagrams. (a) Bulk diagram with low Mn concentration ($10^{-6}$ mol/kg Mn), (b) Bulk diagram with high Mn concentration (1 mol/kg Mn), (c) Surface diagram on rutile MnO₂ with high Mn concentration (1 mol/kg Mn) indicating surface phases 1 = 4O, 2 = 3O, 3 = 2O_b, 4 = O_b + OH_b, 5 = 2OH_b. Additional information in ref.[38] Dashed lines indicate the stability window of water between OER (upper line) and hydrogen evolution (lower line). Panels a and b were produced using materialsproject.org;[39] under CC BY 4.0 license. Panel c was modified from ref.[38] with permission from the Royal Society of Chemistry.

For mechanistic insight, the Tafel slope ($\partial \log i/\partial E$), reaction order with respect to pH ($\partial pH/\partial \log i$) and Nernst slope ($\partial E/\partial pH$) are determined and compared to predictions from the proposed reaction paths.[23,40–42] Unfortunately, the electrochemical parameters are often ambiguous, especially when the conclusions rely on a single mechanistic parameter such as the Tafel slope, since they may be influenced by concomitant processes, e.g., redox transformation of a electrocatalyst component or competing Faradaic reactions, as well as by additional factors including electrical conductivity, blocking of the electrode surface due to gas bubble formation, local pH changes, among others.[43,44] Therefore, complementary spectroscopic investigations as well as reaction product analysis are needed to corroborate the mechanistic insights gained by electrochemical methods.

The assignment of physicochemical processes to the observed features in the CV is often unclear. In addition to redox reactions, adsorption/desorption processes on metallic surfaces can also result in peaks in the CV, where one of the best-known examples is hydrogen underdeposition on Pt.[30] The peak potential of adsorption/desorption processes depends on the concentration of the involved ion in solution, which can be used for their identification.[13] On metal oxides, adsorption/desorption without a redox process is rarely discussed, likely due to localized electronic states.[45–49] Furthermore, different surface facets may have different reaction energies of the same element and thus two anodic peaks do not necessarily indicate redox of two different elements.[50–52] Likewise, one broad peak or shoulder does not necessarily correspond to a single redox reaction but instead may correspond to several redox reactions.[53] This is exemplified in the cathodic peak in Figure 2a, which, in addition to reduction of Mn, comprises contributions from the oxygen reduction reaction (ORR, being the reverse reaction of the OER) as seen by the asymmetry of the anodic and cathodic peaks and the drop of the current below the zero-current baseline (dashed line) at potentials more cathodic than the redox peak. The assignment of a specific redox couple is particularly challenging for multi-metallic oxides and an additional characterization such as XAS should be used to reduce the ambiguity.

A redox transition may also be coupled to a phase transition in an oxide. The expected changes in the thermodynamically stable phases are displayed in E-pH diagrams, also called Pourbaix diagrams. Commonly, the bulk phases are calculated (Figure 3a,b) but for some cases, including the Mn-O-H system, surface phases have also been calculated, e.g., on rutile $\beta$-$MnO_2$ (Figure 3c). The calculations suggest O-terminated $MnO_2$ as the active surface phase at OER conditions (as observed by the overlap of the OER line and the $MnO_2$ phase dominance in Figure 3c). While these plots are helpful to estimate what phases may be observed in an experiment, there are several important aspects that need to be considered but are not included in them. Firstly, all possible reactions are assumed to be in chemical equilibrium and possible kinetic effects are neglected. Secondly, most experimental systems are chemically more complex, e.g., the study of a binary manganese oxide in KOH corresponds to the system Mn-K-O-H, which may stabilize additional phases such as birnessites (e.g., $\delta$-$K_xMnO_{2-y}\cdot zH_2O$) as compared to the system Mn-O-H. The cation concentration (e.g., that of Mn) determines the stability of the phases where low concentrations favour solvated species at otherwise identical pH and E (Figure 3a) and large concentrations favour solids (Figure 3b). Another example is the case of tunnel-structured manganese oxides, whose properties including the framework stability may be strongly influenced by the charge-balancing cations found within their tunnels.[54,55] For instance, when immersed in $Na_2SO_4$ aqueous solution, the structure of cryptomelane-type manganese oxide ($\alpha$-$MnO_2$) collapses due to $Na^+$ intercalation.[56] Furthermore, complexation such as Mn with $PO_4^{3-}$ ions in a phosphate buffer is rarely explicitly included in E-pH diagrams, but can have large effect on the stabilization and destabilization of phases. Calculated E-pH diagrams provide valuable guidance in the interpretation of the active state but additional measurements are required to fully understand possible phase changes occurring during electrochemical experiments.

**Table 1.** Typical analyses of the discussed potential regions

| Potential region | Typically analysed for |
| --- | --- |
| Redox changes | Redox pairs |
| | Number of active sites |
| | Phase changes |
| Double layer | Surface area |
| | Capacitance |
| Electrocatalysis | Electrocatalyst activity, stability, selectivity |
| | Catalytic mechanism |

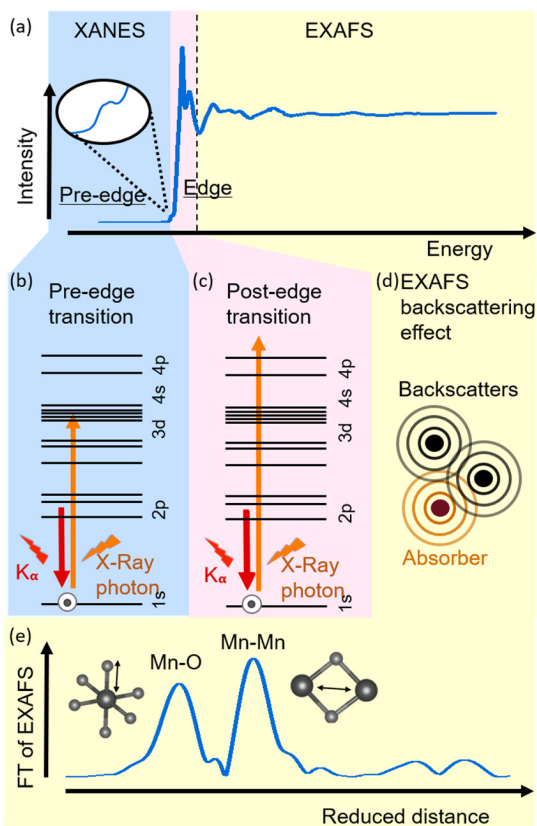

**Figure 4.** Anatomy of an X-ray absorption spectrum and associated physical processes. (a) X-ray absorption spectrum of LiMn$_2$O$_4$ consisting of the XANES (blue and pink highlights) and EXAFS regions (yellow highlight). The inset magnifies the pre-edge. Assignment of features in the spectrum to (b) a pre-edge transition, (c) a main edge transition to the continuum and (d) backscattering and self-interference that produce the EXAFS, which is often displayed as (d) a Fourier transform between 15 and 600 eV where peaks can be assigned to distances between atoms. Dataset in ref. [57]

# Brief fundamentals of X-ray absorption spectroscopy

XAS is an element specific method with high chemical sensitivity that allows distinguishing different valence states of the probed elements and, in some cases, it can provide a deep insight into crystal field splitting and electronic structure. The absorption of an X-ray photon at a specific atom, e.g., Mn, occurs when the energy of the interacting photon is equal or higher than the binding energy of a core electron of the atom. In a typical XAS experiment, the incident photon energy is stepped or swept using a monochromator. The absorption coefficient can either be obtained from the change in transmission through the sample, which is directly recorded (e.g., using ion chambers), using the Beer-Lambert law, or from the fluorescence yield or electron yield of the sample, which is recorded with

suitable detectors in a geometry where these yields are proportional to the absorption coefficient.[58–60]

The quantized nature of the absorption process results in a sharp discontinuity in intensity as a function of energy, which is called the edge. Edges are labeled using the X-ray notation[61] (or IUPAC notation), where capital letters K, L, M and higher are assigned to core holes produced in shells with principal quantum numbers n = 1, 2, 3 and higher, respectively. The spectroscopic notation is also used,[62] which combines the principal quantum number with a small letter, indicating the azimuthal quantum number, where s, p, d, etc. correspond to l = 0, 1, 2, etc., respectively. For example, the Mn-K edge indicates a core hole in a Mn-1s shell, whereas the Mn-L edges indicate a core hole in a Mn-2s or Mn-2p shell. A larger quantum number leads to shorter attenuation lengths in the sample of at least few tens of nanometers (Fig. S1).[63] XAS is most suitable for the analysis of samples around an attenuation length in transmission mode and thinner than an attenuation length in fluorescence mode. Since the attenuation lengths are large as compared to the ionic radii of the top layer (< 0.1 nm for Mn),[64] the study of surface processes requires high-surface area samples so that a large fraction of the measured volume is contributred by the surface. Nanoparticles are an example with high surface to volume ratio, which are also attractive electrocatalytic materials.

**Table 2.** Typical analyses of the discussed spectral regions

| Spectral region | Typically analysed for |
|---|---|
| Pre-edge | Oxidation state<br>Coordination symmetry |
| Main edge | Oxidation state |
| EXAFS | Interatomic distances<br>Number of interactions (coordination number)<br>(Dis)order, i.e., Debye-Waller factor |

Figure 4a shows the X-ray absorption spectrum of a manganese oxide, here LiMnO$_2$, at the Mn-K edge, which can be partitioned into three regions: (i) the pre-edge, (ii) the main edge, and (iii) the extended X-ray absorption fine structure (EXAFS). The spectrum can be described by one electron making a transition to partially occupied states (pre-edge; Figure 4b) or to unoccupied states (main edge; Figure 4c) as well as an electron being ejected from the absorbing atom to the continuum where it scatters between neighboring atoms and interferes with itself (EXAFS; Figure 4d). The pre-edge and edge regions constitute the X-ray absorption near edge structure (XANES), also called near edge X-ray absorption fine structure (NEXAFS).

The pre-edges in transition metal K edges such as the Mn-K edge are assigned to a transition from 1s states to a partially occupied 3d states (Figure 4b), which is forbidden by the dipole selection rule ($\Delta l = \pm 1$). Yet, it can be observed in experimental spectra, often due to mixing of p and d orbitals[65] and in rare cases due to a quadrupole transition.[66] For most transition metal oxides, the pre-edges have low intensity. Notable exceptions are transition metal oxides with low symmetry in the first coordination shell and having low occupation of the d orbitals such as tetrahedral $KMnO_4$ ($d^0$),[65] so that the pre-edge can be used to estimate the coordination symmetry (Table 2). More often, the oxidation state is determined using the pre-edge area or position relative to reference materials (Table 2); the accuracy of the analysis is not discussed in literature. Furthermore, contributions of $e_g$ and $t_{2g}$ states to the pre-edge can be quantified in some cases,[67,68] although the ligand K edges are usually more sensitive to it as compared to the metal K edges.

The main edge is caused by a dipole transition from 1s to unoccupied 4p states or the continuum (Figure 4c). A sharp feature is often observed at the top of the edge, also known as the white line (because it showed up as a white line on photographic plates used in the early days of synchrotron research). The edge position shifts to higher energies with higher oxidation state of the absorbing metal but it also depends on the nature of the metal's ligand. Two related explanations are discussed:[69] firstly, oxidation (donating electrons) lowers the shielding of the core electrons which increases their effective charge so that the energy difference of the transition increases; and secondly, absorber and neighboring scattering ligands with distance R can be treated as a potential well, in which the energy increases as $1/R^2$ (known as Natoli's rule).[70] Usually, a higher formal oxidation state leads to a shorter metal-ligand distance. The empirical determination of the oxidation state by calibration of a series of well-known reference materials is well established. For such a purpose, the energy of the edge must be accurately estimated. Typical definitions of the edge position include the energy at 0.5 intensity in a normalized spectrum, the inflection point, and the area under the edge rise.[60,71] Alternatively, linear combination analysis (LCA) or principle component analysis (PCA) are used.[72–74] The analysis is only valid if the observed oxidation states are a combination of the reference states and no new electronic states emerge. The accuracy of the calibration plot is rarely discussed, yet we have shown calibration of the Mn oxidation state of manganese oxides at the Mn-K edge with an accuracy to about ±0.16 oxidation states[44] (rms error) using the area under the edge. The accuracy of the oxidation state calibration deserves a detailed discussion to be published elsewhere.

The oscillations at energies higher than the main edge are called EXAFS. They are caused by intensity modulations of the absorption coefficient due to electrons that make transitions to the continuum, scatter at neighboring atoms and then interfere with themselves (Figure 4d), forming constructive and deconstructive interference (observed in the spectrum as oscillations). Thus, the EXAFS encodes information about the surroundings of the absorbing atom, including the distance between the absorbing and the scattering atoms, the number of scattering atoms at a given distance and the EXAFS Debye-Waller factor, which relates to the local order (Table 2). More precisely, the most frequent (and not the average) atomic distances are measured, which is an important distinction, e.g., for Jahn-Teller distortions (e.g., of $Mn^{3+}$), i.e., an octahedron with groups of four and two similar distances (Figure 2b). As the EXAFS is not sensitive to bond angles, often the structural insight is discussed in terms of coordination shells. Usually, the EXAFS is isolated from the edge and the energy axis is converted to momentum (wavenumber) space. Subsequently, a Fourier transform is applied which converts the momentum space axis to a real space axis, thus providing the advantage that the coordination shells appear as peaks in the typically used plots of the modulus of the complex EXAFS function (Figure 4e). These plots show the reduced distance on the x-axis, which is smaller than the interatomic distances due to a contribution of the scattering phase in the Fourier transform. The interatomic distances, occupation of a given shell (i.e., coordination number) and Debye-Waller factor can be obtained by fits to the simulated EXAFS. The latter requires scattering factors and phase information that is nowadays most commonly obtained from first principles calculations performed on a suitable atomic model. The interatomic distances from optimized fits can be as accurate as ±0.01 Å for shorter distances, e.g. metal-oxygen, while the shell occupation is only accurate to about ±1 due to the strong correlation with the Debye-Waller factor.

In our Fourier transform example (Figure 4e), two peaks are prominently visible, where the one at lower distance has been assigned to the Mn-O shell and the one at higher distance has been assigned to a Mn-Mn shell. Often, these assignments can be guessed with some experience based on the reduced distance but it is prudent to confirm the assignment using simulations. The EXAFS of single scattering events is usually observed to about 5 Å of the absorbing atom,[60,69,75] however, positive interference of multiple scattering events can extend it to about 8 Å (Figure S2) for metal oxides with long-range order. This means that analysis of the EXAFS does not need extended long-range order and in fact, even a single shell, e.g., the hydration sphere of solvated ions, can be investigated.

The analysis of the Mn-L edges (and other first row transition metal L edges) can be performed similarly with some important differences. The involved transitions cannot be treated as a one-electron process as the wave functions of the core and valence states overlap.[76] Therefore, the Mn-L edge corresponds to the system density of states (as opposed to the electron density of states measured at the Mn-K edge) and the overlap gives rise to multiplett effects in the pre-edges. Consequently, the pre-edges are much larger in Mn-L edge spectra and the main edge is barely visible. Having pre-edge character, the Mn-L edges depend strongly on the coordination symmetry and also on the number of d electrons, i.e., the oxidation state. It is often determined empirically by calibration of the pre-edge area or peak to known reference materials. Alternatively, PCA and LCA are also used for L edge analysis. The EXAFS is also less visible at the Mn-L edges. It is rarely recorded and requires special efforts in the analysis as a very limited energy range can be used in addition to exhibiting weak signals.

In summary, the average metal oxidation state is obtained from analysis of the XANES, while structural insights of the most frequent motifs are obtained from analysis of the EXAFS, even for samples without long-range order. Since the attenuation length of the incoming (and outgoing photons) is much lower in water as compared to Mn oxides (Fig. S1), it is well suited for in situ measurements in thin layers of aqueous solutions. Insight into the electronic and geometric structure in a single experiment makes XAS an ideal complementary method to resolve ambiguities in electrochemical experiments.

# When and how do electrocatalysts change?

What can alter a freshly synthesized sample? In the worst case, simply exposure to our atmosphere as well as any subsequent sample preparation step and experiment performed. We have summarized the possible changes and when they may occur in Table 3. The processes leading to changes and examples are discussed in the following subsections. They are presented according to the moment when these are investigated with respect to the catalytic process: before (pre-catalysis investigations), during (in situ and operando investigations) and after (post-mortem investigations) catalysis.

**Table 3.** Possible changes of as-prepared samples before, during and after electrochemistry.

| When does the electrocatalyst change? | Nature of change |
| --- | --- |
| During electrode preparation and storage | Chemical reaction |
| Immersion of electrodes in electrolyte | Chemical reaction |
| In electrolyte under polarization* | Electrochemical reaction |
| During/after catalysis ** | Chemical & electrochemical reaction |

\* Exposed to non-catalytic potentials
\*\*Exposed to catalytic potentials

**Pre-catalysis investigations**

The surfaces of many transition metal oxides, including common Mn oxides, react with their environment after synthesis, which could be due to exposure to oxygen in our atmosphere, to solvents and/or chemicals during electrode preparation, e.g., when preparing an electrocatalyst ink from powders, or to the electrolyte when the transition metal oxides are immersed and thus mounted in the electrochemical cell (Figure 5). As these changes are most pronounced near the electrocatalyst surface, we focus on Mn-L edge XAS where the near surface region contributes more to the signal as compared to Mn-K edges.

An example of a surface reaction due to storage is given by rocksalt MnO, where Mn is expected to be in oxidation state 2+. A typical spectrum of $Mn^{2+}$ has a main peak at 641.1 eV with two minor peaks at 640.0 eV and 642.4 eV as found in the Mn-$L_3$ edge spectrum of $MnSO_4 \cdot 4H_2O$. Note that while the absolute energy values depend on the energy calibration, which is not harmonized in the field, the relative energy values and the fingerprint are still a valid diagnostic. Yet, the Mn-$L_3$ spectrum of "MnO" looks drastically different to that of $MnSO_4 \cdot 4H_2O$, and is composed of a superposition of the peaks of $Mn^{3+}$ oxides (such as $Mn_2O_3$) and $Mn^{2+}$ oxides (such as $MnSO_4 \cdot 4H_2O$), as shown in Figure 5a. The presence of the peaks from both oxidation states suggests the partial oxidation of the near surface. In mild cases of oxidation, the Mn-K edge looks as expected for $Mn^{2+}$ since oxidation only takes place at the electrocatalyst surface. However, in severe cases, oxidation may also lead to changes in the Mn-K edge. An example for this are metallic Mn foils that not only show the spectrum of $Mn^{2+}$ in Mn-L edge XANES, but also display an additional shoulder in the Mn-K edge

XANES. Thus, we recommend to check physicochemical properties of the samples periodically, particularly for long storage times, and ideally before an electrochemical measurement.

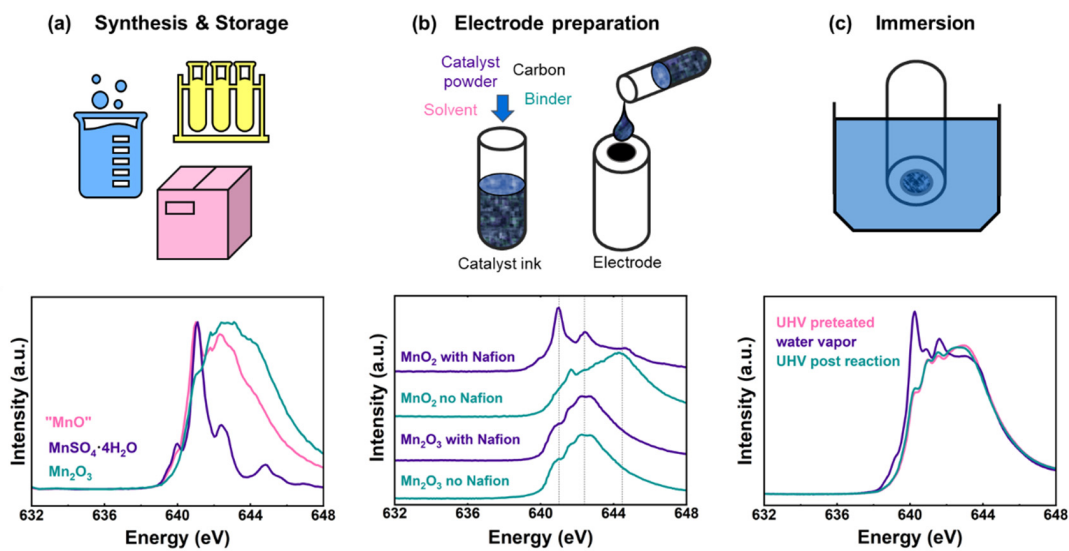

**Figure 5.** Possible changes that electrocatalysts undergo due to (a) storage, (b) electrode preparation and (c) immersion in an electrolyte. (a) Mn-$L_3$ edge of nominally MnO that shows unexpected oxidation. Mn-$L_3$ spectra of MnSO$_4$·4H$_2$O and Mn$_2$O$_3$ are shown for comparison. They were recorded at the SGM beamline at Canadian Lightsource (CLS) in electron yield mode. (b) Mn-$L_{2,3}$ edges of electrode films containing Mn oxide with and without Nafion showing reduction of Mn$^{4+}$ in the presence of Nafion. (c) Mn-$L_3$ edge of a Mn-containing perovskite oxide showing reduction in water vapor. Panel (a) was modified from ref.[77] with permission from the ACS, panel (b) is adapted under CC BY 4.0 licence from ref.[77] and panel (c) is adapted under CC BY 3.0 license from ref. [78] Dataset in ref. [57]

Ink-casting is a popular method to prepare electrodes for electrocatalytic investigations. The catalyst ink is typically prepared by mixing the oxide particles with a suitable dispersion medium such as an alcohol-water mixture or tetrahydrofuran (THF). Usually, further additives are added, namely high surface area carbon to enhance the electrode conductivity and a binder, typically an ionomer, to ensure mechanical stability of the electrode film. However, when preparing catalyst inks, some considerations are needed but often overlooked. Some transition metal oxides react with the dispersing solvent,[79] in which case a different solvent should be used. Carbon may accelerate the amorphization of some oxide surfaces.[80] Additionally, carbon may oxidize electrochemically to CO$_2$ at similar potentials as the OER and can act as a co-catalyst,[81,82] e.g., during oxygen conversion, so that carbon may affect the product currents in addition to inducing changes of the electrocatalyst material.[83]

Our example for Mn oxides is the reaction with the very popular ionomer binder Nafion (Figure 5b).[77] Mn$^{4+}$O$_2$ reacted chemically with Nafion to form Mn$^{2+}$, while (Mn$^{3+}$)$_2$O$_3$ was unaffected by the addition of Nafion. A similar trend was seen for other Mn$^{4+}$-containing oxides. We speculated that Mn$^{4+}$ interacts with the electron donor groups in the binder.[77] It is currently unclear whether other highly oxidized transition metals may also undergo redox reactions with Nafion or whether other ionomers react similarly. Consequently, we recommend to check for physicochemical changes after ink preparation.

The as-prepared electrodes may further react prior to the desired electrochemical experiment due to immersion in the electrolyte. Many transition metal oxides are prompt to react chemically with water (being necessary for the OER). For example, Pr$_{0.2}$Ca$_{0.8}$MnO$_3^{3.8+}$O$_3$ reacted with water vapor to produce a Mn$^{2+}$ species (Figure 5c). Other transition metal oxides, in particular those containing Ni, are known to incorporate Fe from alkaline electrolytes, which strongly modifies their electronic and thus catalytic properties.[84–87] Also, in the absence of Fe, exposure to hydroxide solutions may change the surface and even the bulk phase of transition metal oxides after sufficiently long exposure.[88] Finally, elements may dissolve from the electrocatalyst into the electrolyte changing the composition and thereby other properties, such as redox potentials or the catalytic reaction's overpotential.[89–91] Dosaev et al.[92] recently studied Mn-bases spinels as synthesized, in the ink suspension and after soaking in hydroxide electrolyte, which oxidized Mn$_3$O$_4$ but not MgMn$_2$O$_4$. We recommend similar control experiments of the electrode soaked for an extended time (at least for the same duration of the intended experiment duration, but ideally much longer) to elucidate possible changes prior to electrochemical experiments.

**Post-mortem investigations**

To investigate how the state of an electrocatalyst changes due to applied potential, i.e., due to electrochemical reactions, it is common to resort to post-mortem experiments as some electrochemical changes can be resolved in this kind of investigations, namely those that result in the formation of stable phases. XAS is particularly useful for this purpose, as it allows to elucidate phase changes even in cases where amorphization takes place, which is not the case for techniques in which crystallinity is a prerequisite, e.g., diffraction-based techniques.

The composition may change electrochemically without significantly affecting the geometric structure. This is a typical scenario for charging or discharging the bulk of a battery material but it is less discussed in the field of electrocatalysis, yet, several typical battery materials such as $LiCoO_2$, $LiMPO_4$ (M = Mn, Fe, Co) and $LiMn_2O_4$ have been investigated as electrocatalysts for the OER.[40,93–97] In our example, we studied the OER on $Li_xMn_2O_4$ (Figure 6a,b) using a rotating-ring disk electrode setup where the ring electrode qualitatively detected the oxygen produced at the disk electrode.[94] The redox potential for delithiation, i.e., $Li_1Mn^{3.5+}_2O_4 \rightarrow Li_0Mn^{4+}_2O_4$, can be calculated as it depends on the electrolyte composition (i.e., pH and Li concentration).[94] The onset of the OER taking place at the disk electrode (defined therein as the potential at which 5 µA were reached at the ring electrode) shifted to more anodic potentials upon decreasing the pH, translating into lower OER activity, with higher potential for delithiation (Figure 6a). Baumgung et al.[94] prepared samples for post-mortem XANES analysis which showed the expected oxidation (of the bulk) as an edge shifts to higher photon energy for high delithiation overpotential (labeled pH 12 in Figure 6b). As a control experiment, a sample where the delithiation is not expected to occur also did not show an edge shift (labelled pH 14 in Figure 6b). We conclude that the results agree with other literature reports[98–100] where an oxidation of the metal site approaching $Mn^{4+}$ reduces the activity for the OER.

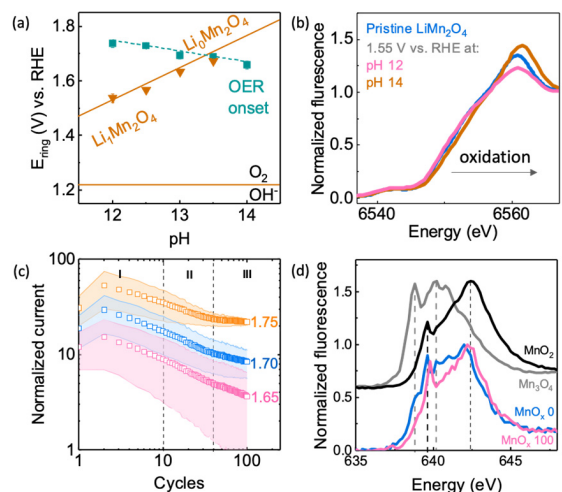

**Figure 6.** Post-mortem investigations of electrocatalyst changes by XAS. (a) Delithiation potential (orange line) and OER onset (green line) of $Li_xMn_2O_4$ and (b) Mn-K edge pristine and post-mortem after conducting the OER at pH 12 and pH 14. (c) Trend of normalized current with cycling at 3 selected potentials of an electrodeposited Mn oxide film, and (d) comparison between the film before voltammetric cycling (blue) and after recording 100 cycles (pink). Panels (a), (b) have been modified from ref.[94] under CC BY 4.0 license and panels (c) and (d) have been modified from ref.[43] under CC BY 4.0 license.

Conditioning by potential cycling is a common procedure to activate catalysts prior to electrocatalytic studies.[91,101–104] The electronic and geometric structure of the activated electrocatalyst often differs from the as-prepared electrocatalyst. As an example, we investigated the change in normalized current with voltammetric cycling for electrodeposited Mn oxide (Figure 6c,d).[43] At 1.75 V vs. RHE (a potential at which the OER takes place), a constant current is reached after about 40 cycles, while at less anodic potentials it decreases continuously upon cycling. While no significant changes were found in the Mn-K edge, the Mn-$L_3$ edge showed clear spectral changes that suggested oxidation of the near-surface toward $Mn^{4+}$ by comparison of the activated electrocatalyst (100 cycles) with a spectrum of $Mn^{4+}O_2$ (Figure 6d). We concluded that the activation procedure led to oxidation of the electrocatalysts, with the changes being restricted to near the surface where catalysis occurs.[43] Similar post-mortem studies after electrocatalysis are commonly performed to elucidate the nature of the activated catalysts.[91]

**In situ investigations**

In situ or even operando experiments are necessary to understand both reversible and irreversible changes of the electrocatalyst as well as reactive transient states (if the time resolution is appropriate). Any combined EC and XAS experiment on transition metal oxides for studying the OER at room temperature is an in situ spectroscopic experiment as they need to be conducted in an electrochemical cell including electrolyte and electrodes. In our definition, the reaction (e.g., the OER) must be taking place during a spectroscopic operando experiment[105] and this needs to be proven by a qualitative or quantitative measurement of the product. For an electrocatalytic study, this could be the measured current if the Faradaic efficiency is known, or ideally, a simultaneous direct or online detection of the catalytic product. Thus, we use a stricter but clear definition of an electrocatalytic operando experiment as compared to the often found and somewhat ambiguous "electrocatalyst under reaction conditions".

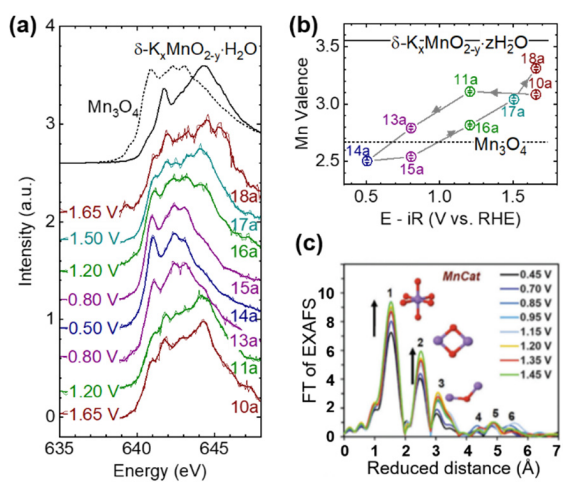

**Figure 7.** In situ investigations of electrocatalyst changes with applied potential. (a) Mn-L$_3$ edge spectra of electrodeposited MnOx in 0.1 M KOH cycled from 1.65 V to 0.80 V and back to 1.65 V vs. RHE. (b) Extracted valence changes with applied potential showing hysteresis. (c) Mn-K edge EXAFS of electrodeposited MnOx (MnCat) in 0.1 M KPi (pH 7) for potentials between 0.45 V to 1.45 V vs. NHE (0.04 to 1.04 V vs. RHE). Panels (a) and (b) are reproduced from ref.[18] with permission from the ACS. Further permission related to the these figures should be directed to the ACS. Panel (c) is reproduced from ref.[106] with permission from the Royal Society of Chemistry.

As expected from the E-pH diagrams (Figure 3), transition metal oxides such as manganese oxide are oxidized and may undergo changes in their structural features under reaction conditions. Our example follows the changes occurring in the Mn-L$_3$ edge XAS of electrodeposited MnO$_x$ from 1.65 V vs. RHE, a potential where the OER takes place, to 0.80 V vs. RHE, a potential where the ORR takes place, and back to 1.65 V vs. RHE (Figure 7a). The example also illustrates finer voltage resolution as compared to post-mortem studies, which mainly compare before and after some known process such as an electrocatalytic reaction or a redox process. The spectra in Figure 7a clearly displayed reversible changes between that of the δ-K$_x$MnO$_2$ reference (birnessite) recorded at 1.5 and 1.65 V vs. RHE and that of Mn$_3$O$_4$ (spinel) recorded at 0.5 and 0.8 V vs. RHE. Note that the spectra are susceptible to bubble formation during the OER at high potentials (e.g., spectra 18a). The spectra collected at 1.2 V vs. RHE look similar to either that of birnessite or spinel depending on which was measured before. This indicates hysteresis due to slow kinetics of the phase changes and highlights the importance of the history of the electrode. The hysteresis can also be seen clearly when the oxidation state is calculated by calibration of the centroid under the Mn-L$_3$ edge with respect to reference materials (Figure 7b). Hysteresis of the Mn oxidation state with applied potential is known in the electrochemical capacitor community[107] but rarely discussed in the electrocatalysis field despite the popularity of cyclic voltammetry for electrocatalyst investigations. Regardless of the hysteresis, the analysis shows that Mn reduction precedes the ORR, while Mn oxidation precedes the OER. Thus, both active states differ markedly from the as-prepared oxide.

Changes in the oxidation state can also be resolved by Mn-K edge XAS for samples with high surface to mass ratio such as porous films[106] or nanoparticles.[108] In situ XAS is especially powerful when metastable phases need to be studied such as the nucleation of MnO$_2$.[109] Previously unknown metastable phases have also been identified such as shown for α-CoO$_2$H$_{1.5}$·0.5H$_2$O using PCA in a recent report, which could not be prepared for post-mortem investigations as they only exist in situ.[110] When transition metal oxides undergo redox reactions, their interatomic distances change accordingly. As discussed earlier, a change in edge position, and thus oxidation state, is expected to be proportional to 1/R$^2$ so that higher oxidation leads to shorter M-O bonds. This must also lead to the shortening of other interatomic distances, e.g., M-M, and/or other changes to the structure of solids, e.g., the connectivity of the coordination polyhedra. EXAFS analysis is ideally suited to resolve these changes. In our next example of an electrodeposited Mn oxide,[100] the coordination number of Mn-O in the first coordination shell changed as well as the number of Mn-Mn interactions (Figure 7c), which was interpreted as changes in the number of μ-oxo bridges in the oxide, i.e., a change in polyhedron connectivity. Similar changes were also reported for electrodeposited Co oxide.[111]

In summary, operando and in situ XAS is mainly performed to elucidate the expected electronic and structural changes with applied potential. Yet, the unexpected changes such as hysteresis in cyclic experiments or metastable intermediate phases show the true potential of combing XAS and electrochemical experiments.

## Summary & perspective

In this Feature Article, we reviewed what XAS can tell us about the active state of Mn oxides during the OER with references to other first row transition metal oxides. In the introduction, we highlighted that the macroscopic insight from electrocatalysis can be combined with the atomistic insight from XAS, such as the local coordination environment or polyhedral connectivity. The product yield in electrocatalysis creates a flow rate of electron, i.e., an electric current. XAS measures the electrons in the outer shell of atoms, i.e., the oxidation state of the metal, so that the two can be correlated. In order to put the desired correlations on a sound scientific basis, we briefly introduced the fundamental of electrochemistry where we discussed the key processes of redox changes, double layer charging and electrocatalysis, all of which may contribute to the measured currents. E-pH diagrams, also called Pourbaix diagrams, were introduced, highlighting their relevance for in situ XAS experiments as they can provide some guidance about what changes to expect for a given combination of pH and applied potential. We continued to discuss the X-ray absorption processes that create the analyzed features in XAS, namely the edge shift and EXAFS analysis of interatomic distances as well as number of scattering atoms. Based on this foundation, we discussed how XAS was used to identify electrocatalyst changes of the oxidation state due to storage, electrode preparation and immersion in an electrolyte. Irreversible changes of the oxidation state and local coordination environment during electrocatalyst activation can be detected by post-mortem XAS, while in situ XAS allows finer voltage (or time) resolution to follow expected changes in current studies. Moreover, in situ XAS enables the study of metastable phases, as well as the investigation of reversible and reactive processes. In the following paragraphs, we will give a perspective on the operando XAS studies needed to close current knowledge gaps.

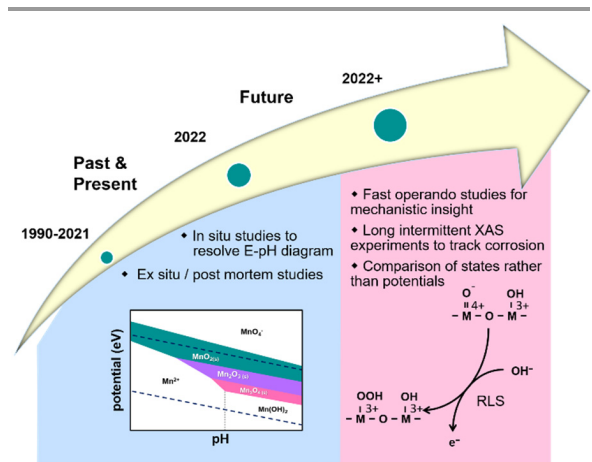

**Figure 8.** Possible evolution of combined electrochemical and XAS investigations. The inset was produced using materialsproject.org;[39] under CC BY 4.0 license.

The relevant time scales of solid state and electrocatalytic processes differ drastically and both are not matched by current XAS studies. Current conventional in situ XAS studies are performed with durations of a few minutes to an hour per spectrum, which is appropriate to study the faster changes in E-pH diagrams. While scan durations of the order of minutes are sufficiently fast to study slow processes such as corrosion and bulk phase changes, the expected changes may occur on time scales longer than typical synchrotron experiments of one to two weeks, which is not compatible with usual synchrotron operation where users apply for an allotment of one or few measurement weeks. A possible solution might be implementing alternative operation modes where one applies for an allotment of a few hours each week over a long duration. Another solution could be the use of lab-based XAS systems, albeit the cost of lower X-ray intensity. It will be crucial for the optimization of electrocatalytic materials with relevance for practical applications to understand the origins of degradation that often occur on time scales much longer than those currently investigated in most academic reports.

The currently used XAS scan durations are also too slow to resolve changes with relevance to the catalytic mechanism such as Mn redox changes of the order of tens to hundreds of milliseconds for electrodeposited Mn oxide.[100] Quick EXAFS (QEXAFS) performed with a special monochromator reduces the scan time for a spectrum to seconds or even milliseconds.[112,113] However, such scan time is too fast for the acquisition time of most fluorescence detectors. Thus, these experiments are performed in transmission mode, which less common for electrocatalytic studies and technically challenging at metal L edges due to the absorption of water.[114] Another approach to faster XAS is implemented using conventional monochromators fixed to a well-selected excitation energy and measuring the change in intensity using appropriate detectors including fluorescence detectors. The method has been recently reviewed by Tesch and Simonov[115] in the context of electrocatalytic reactions. It is commonly used to track the X-ray signal during cyclic voltammetry but can also be used to record transients during potentiostatic measurements.[116] These faster XAS measurements will be crucial to separate changes of the oxidation state due to non-catalytic from those occurring during catalysis, and for precise identification of the redox states involved in the electrocatalysis of the OER and of other reactions.

The final point that can advance future combined EC and XAS experiments is the study of states rather than applied potentials. Relevant states are the redox changes of transition metals such as Mn. They may be identified by electrochemical features in cyclic voltammograms (Figure 2a) but especially Mn oxides have redox transitions that do not result in obvious voltammetric features.[53] These redox transitions can be resolved using spectroscopic methods, e.g., UV-Vis spectroscopy[117,118] and XAS,[18,119–121] which have signals proportional to the oxidation state (a charge, Q). The time derivative of the signals (dQ/dt) is thus the redox current.[111] Alternatively, the redox couples can be identified by fitting the oxidation state from XAS to (modified) Nernst equations.[111] The nature of the redox couples directly relates to an atomistic understanding of the key steps of the catalytic mechanism, e.g., of the rate-limiting step (RLS) as shown in Figure 8, and their midpoint potential is crucial to understand charge transfer from/to semiconductors, e.g., in the framework of the Marcus-Gerischer theory.[122–124] Identification of the relevant redox couples is thus a natural choice for mechanistic discussions and has led to the recent insight that the OER is a first order reaction with respect to the Mn density of states, while the ORR is of second order.[125] Moreover, the $Mn^{3+/4+}$ redox couple is essential for the evolution of oxygen in natural photosynthesis as well as for electrodeposited Mn oxides[126] because $Mn^{4+}$ togetherwith $Mn^{3+}$ has been proposed as a necessity for the OER and because its midpoint potential is similar to that of the OER. The $M^{3+/4+}$ redox couple is likewise important for the OER on other transition metal oxides,[126] where the detection of $M^{4+}$ has been correlated with oxygen evolution for a Co oxide.[127] Moreover, the midpoint potential of the $Mn^{3+/4+}$ redox couple also was used to rationalize electrocatalytic trends of the also ORR.[128] We expect that combined operando EC-XAS studies willcontinue to unravel important mechanistic details of the OER and of other reactions as well as provide crucial physical insight to built improved models for the knowledge-guided design of electrocatalysts.

## Author Contributions

MR: conceptualization, data/literature curation, funding acquisition, project administration, resources, supervision, writing; DMM: visualization, writing; JV: visualization, writing; DA: data/literature curation, writing.

## Conflicts of interest

There are no conflicts to declare.

## Data availability statement

Previously unpublished data for this paper are available at Figshare at https://doi.org/10.6084/m9.figshare.20393064. For other data refer to the original publication.

## Acknowledgements

Part of the research described in this paper was performed at the Canadian Light Source, a national research facility of the University of Saskatchewan, which is supported by the Canada Foundation for Innovation (CFI), the Natural Sciences and Engineering Research Council (NSERC), the National Research Council (NRC), the Canadian Institutes of Health Research (CIHR), the Government of Saskatchewan, and the University of Saskatchewan. Some experiments were performed at the CLÆSS beamline at ALBA Synchrotron with the collaboration of ALBA staff. We thank the Helmholtz-Zentrum Berlin für Materialien und Energie for the allocation of synchrotron radiation beamtime. This project has received funding from the European Research Council (ERC) under the European Union's Horizon 2020 research and innovation programme under grant agreement No. 804092.

## Notes and references

# What X-ray absorption spectroscopy can tell us about the active state of earth-abundant electrocatalysts for the oxygen evolution reaction

Marcel Risch,*[a] Dulce M. Morales,[a] Javier Villalobos[a] and Denis Antipin[a]

*Nachwuchsgruppe Gestaltung des Sauerstoffentwicklungsmechanismus, Helmholtz-Zentrum Berlin für Materialien und Energie GmbH, Hahn-Meitner-Platz 1, Berlin 14109, Germany.*
* marcel.risch@helmholtz-berlin.de.

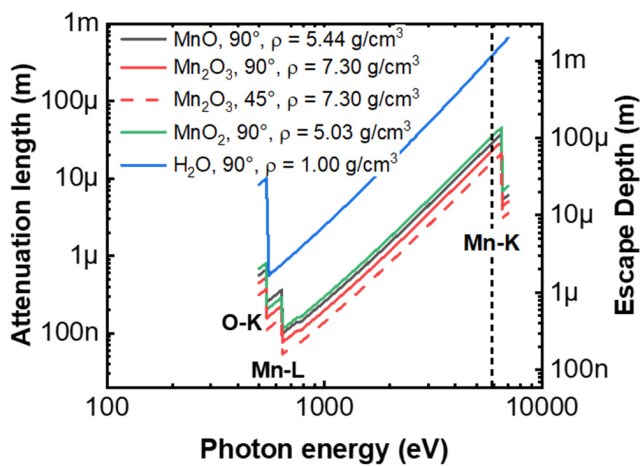

**Figure S1.** Calculated attenuation lengths of the Mn oxides MnO, $Mn_2O_3$ and $MnO_2$ as well as water. The calculation was performed using the Henke Tables and densities were estimated. 90° incidence is typical for transmission measurements while 45° incidence is typical for fluorescence measurements. The dashed line indicates the Mn-Kα line, which is used to calculate the attenuation length in fluoresnce experiments (at the Mn-K edge). We estimated the escape depth as trice the attenuation length (i.e., 99.7 % attenuation). Dataset in ref. [57]

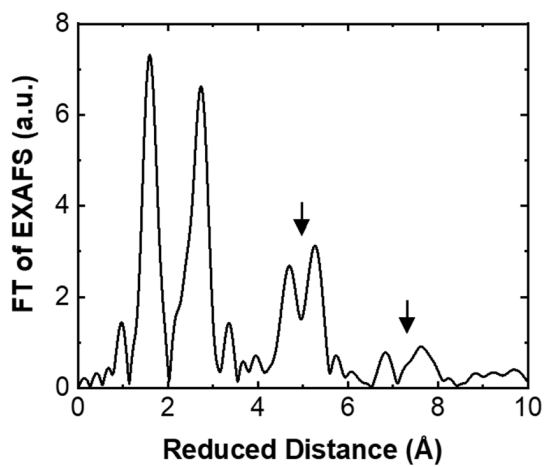

**Figure S2.** Fourtier Transform (FT) of the EXAFS of commercial $LiMnO_2$ (Sigma Aldrich) measured in transmission mode at the ALBA synchrotron. For the measurement, a pellet of $LiMnO_2$ diluted with boron nitride (5 wt%) was pressed and encapsulated in Kapton foil. The arrows indicate FT peaks due to constructive interference in the orded crystal structure. Dataset in ref. [57]